\documentclass[doublecol,cite]{epl2} 
\usepackage{graphicx,epsfig}
\usepackage{subfigure}  %apparently this package is obsolete (but it works). the new one is subfig, but it doesn't work.
\usepackage{amsmath}
\usepackage{amsfonts}
\usepackage{amssymb}
\usepackage{xfrac}
\usepackage{enumitem}
\usepackage{xcolor}
\usepackage{color}
\usepackage{epstopdf}
\usepackage{ulem}
\usepackage[hidelinks]{hyperref}
\usepackage[capitalise]{cleveref}
\usepackage{bbold}
\usepackage{fancyhdr}

\usepackage{bbding}
\usepackage{amssymb}
\newcommand{\1}{\begin{equation}}
\newcommand{\2}{\end{equation}}
\newcommand{\ea}{\begin{eqnarray}} 
\newcommand{\ee}{\end{eqnarray}}
\newcommand{\4}[2]{{\frac{#1}{#2}}}
 
\newcommand{\Int}[2]{{\int\limits_{#1}^{#2}}}

\newcommand{\de} {{\rm d}}

\newcommand{\vsc}{\vskip .01cm \noindent}

\title{Optimal Active Particle Navigation meets Machine Learning}

\date{\today}

\shorttitle{Optimal Active Particle Navigation} %Insert here a short version of the title if it exceeds 70 characters

\author{Mahdi Nasiri\inst{1} \and Hartmut L\"owen\inst{2} \and Benno Liebchen\inst{1}\thanks{E-mail: \email{benno.liebchen@pkm.tu-darmstadt.de}}}
\shortauthor{Mahdi Nasiri \etal}

\institute{                    
  \inst{1} Institute of Condensed Matter Physics, Technische Universität Darmstadt - D-64289 Darmstadt, Germany\\
  \inst{2} Institut für Theoretische Physik II: Weiche Materie, Heinrich-Heine-Universität Düsseldorf, D-40225 Düsseldorf, Germany}

\pacs{82.70.Dd}{Colloids}
\pacs{05.40.Jc}{Brownian motion}
\abstract{
The question of how ``smart'' active agents, like insects, microorganisms, or future colloidal robots need to steer to optimally reach or discover a target, such as
an odor source, food, or a cancer cell in a complex environment has recently attracted great interest. 
Here, we provide an overview of recent developments, regarding such optimal navigation problems, from the micro- to the macroscale, and give a perspective by discussing some of the challenges which are ahead of us. Besides exemplifying an elementary approach to optimal navigation problems, the article focuses 
on works utilizing machine learning-based methods. Such learning-based approaches can uncover highly efficient navigation strategies even for problems that involve e.g. chaotic, high-dimensional, or unknown environments
and are hardly solvable based on 
conventional analytical or simulation methods.}

\begin{document}
\maketitle

\section{Introduction}
%\underline{Optimal navigation at the macroscale --}
Before the start of an airplane, the conductor often runs a software to plan a path to a target destination. This path aims to represent the best possible compromise between traveling time, fuel consumption, and safety aspects for given environmental conditions, such as the current wind and temperature pattern, as well as airspace occupancy.
Similar optimal navigation problems where an active (self-propelled) agent, which can control its direction of motion, or its speed, in a complex environment also occur for robots and autonomously driving cars. These agents are commonly equipped with cameras, processing units and actuators \cite{smith2011persistent, levinson2011towards,galceran2013survey}, allowing them to locally perceive their environment and to use it to achieve a goal such as completing a race track as fast as possible or to efficiently collect waste. 
Here, the cameras typically create images of the environment, that are fed into the processing unit, which then translates the received information into a desired action (motion command) and sends it to an actuator, which initiates the motion. In the animal kingdom, the ability to develop efficient navigation strategies can assist the survival of species. To reach their breeding grounds, some turtles, for example, have to find an efficient path through the ocean over hundreds of kilometers \cite{lohmann2004geomagnetic, sequeira2020animal}. Similarly, insects need strategies to find odor sources by navigating through complex molecular patterns that can even be chaotic due to turbulent air streams \cite{klimontovich1991turbulent, carde2008navigational, bau2015modeling, rigolli2022alternation}.\\
\underline{Smart microswimmers --}
Besides macroscale agents, even microorganisms 
can perceive information from their environment and use it for navigation. 
For example, the ability of sperm cells to sense gradients in the concentration of those chemicals which are emitted by the egg cell \cite{spehr2003identification, eisenbach2006sperm} is crucial for the survival of many species. Similarly, bacteria possess a remarkable spectrum of biochemical sensors which allow them e.g. to measure gradients in oxygen, nutrient, or autoinducer concentration \cite{berg2004coli,bi2018stimulus,laganenka2016chemotaxis} and to use them for navigation. 
\\
Besides biological microswimmers, since almost two decades \cite{paxton2004catalytic} also synthetic microswimmers have become available, and are currently studied together in the research field of active matter \cite{elgeti2015,cates2015b,bialke2015,menzel2015,klapp2016,ebbens2016,hagan2016,maass2016,patteson2016,
zottl2016,katuri2016,moran2017,mallory2018,niu2018,julicher2018,bar2019,agostinelli2020micromotility,bechinger2016active,hecht2021introduction,liebchen2021interactions}. Synthetic microswimmers can be steered by external 
fields \cite{Chen2014,Mano2017,Palacci2013, sanchez2015chemically,dai2016programmable,dong2016highly,xuan2016near, Driscoll2017,Martinez-Pedrero2017,Nedev2015,Zong2015,Moyses2016,liu2016self,Liu2018,Mousavi2019,Demirors2018,Lozano2019,jahanshahi2020realization}
(or even with feedback control systems ~\cite{Khadka2018,fernandez2020feedback,lavergne2019group}) and can react to their environment through various forms of taxis \cite{liebchen2018synthetic,tsang2020roads}, which may be used in the future to help them navigate through our blood vessels to detect and perhaps repair mutated cells \cite{harari2016,bunea2020recent}, transport drugs to cancer cells \cite{patra2013intelligent,ceylan20193d,alapan2018soft} or perform microsurgery \cite{vyskocil2020cancer}.

\noindent While optimal navigation problems at the macroscale have been studied for decades \cite{hart1968formal, gasparetto2015path}, based on methods such as optimal control theory and dynamic programming \cite{martin2001optimal,techy2009minimum, lewis2017optimal, mohamed2018optimal}, and more recently reinforcement learning \cite{zhang2015geometric, xin2017application}, at the microscale corresponding explorations have started only recently.
% Following the wide-spread importance of optimal navigation problems for self-propelled agents like airplanes, fish and turtles at the macroscale a broad set of powerful analytical and numerical methods has been developed in the physics, mathematics, and engineering literature . These include various methods from optimal control theory and dynamic programming \cite{martin2001optimal,techy2009minimum, lewis2017optimal, mohamed2018optimal} as well as methods based on reinforcement learning \cite{zhang2015geometric, xin2017application}. 
%One recent example has been studied by part of the DeepMind team in ref. \cite{mirowski2016learning}, which uses 
%(deep) reinforcement learning (A3C) for 
%optimal end-to-end navigation of agents which have a certain field of view to observe their environment and
%the resulting RBG image, encoded with a convolutional network, to 
%make navigational decisions.
% In contrast, while active particles at the microscale, or microswimmers, have been intensively discussed recently (for recent reviews, see \cite{elgeti2015,cates2015b,bialke2015,menzel2015,klapp2016,ebbens2016,hagan2016,maass2016,patteson2016,
% zottl2016,katuri2016,moran2017,mallory2018,niu2018,julicher2018,bar2019,agostinelli2020micromotility,bechinger2016active,hecht2021introduction,liebchen2021interactions}), the exploration of optimal navigation problems for such particles has started only very recently. 
Here, the smallness of the particles leads to various new challenges: (i) Microswimmers are subject to significant fluctuations due to Brownian motion (or errors and delays in the steering protocol), hence they cannot accurately predict the outcome of their navigational maneuvers. (ii) Microswimmers interact hydrodynamically with walls, obstacles, and other microswimmers, which can qualitatively change the required navigation strategy to reach a target fastest \cite{daddi2021hydrodynamics}. (iii) The displacement rate of microswimmers due to their environment can exceed their self-propulsion speed (typically $\sim \mu m/s$) by orders of magnitude, e.g. in the blood vessels which is opposite e.g. to airplanes in the wind. (iv) For many prospective applications, microswimmers will face unknown environments with only local information about their surroundings available and will require transferable navigation strategies.
\vsc
\section{Optimal point-to-point navigation} 
\begin{figure*}
\begin{center}
\includegraphics[width=0.99\textwidth]{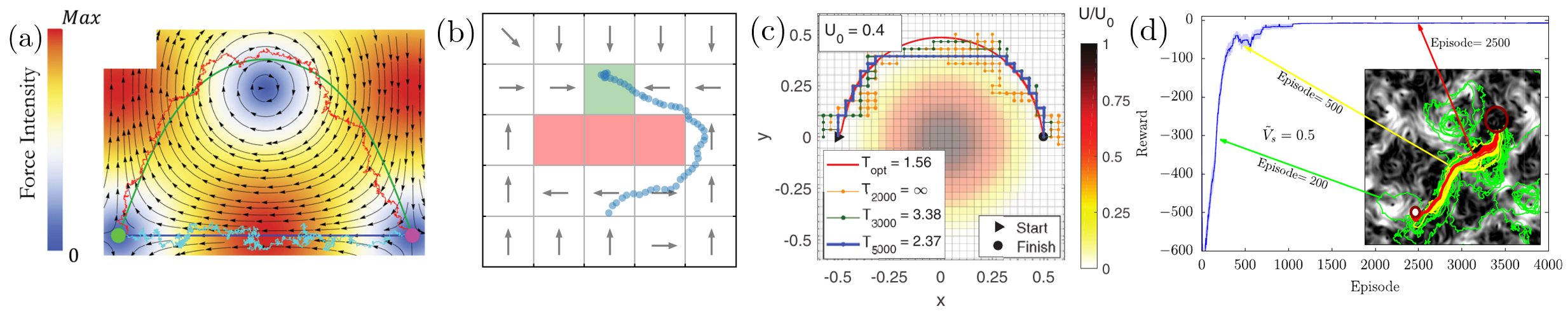}
\caption{\small Optimal active particle navigation. (a) Trajectories (red and cyan) of active particles in the presence of Brownian noise. The particles steer such that they try to follow the optimal (green curve) and the shortest path (blue line) of the underlying determinstic problem in Taylor-Green flow (arrows and background color) \cite{piro2022optimal}. (b) Snapshot of a learned trajectory (blue curve) and corresponding policy map (arrows) in presence of virtual obstacles (red cells) in experiments with feedback-controlled colloids moving toward a goal (green cell) \cite{muinos2018}. (c) Trajectories of an active particle obtained by Q-learning for point-to-point navigation in a mexican hat potential (background color) after 2000 (yellow), 3000 (green) and 5000 (blue) training episodes in comparison with the exact optimal trajectory (red) \cite{schneider2019}. (d) Time-evolution of the total reward (blue curve) and exemplaric trajectories (inset) for active particles in turbulent flows \cite{biferale2019}. See refs for more details.}
%Optimal active particle navigation. (a) Stochastic trajectories (red and cyan paths) %of the active particles following the underlying deterministic time optimal (green %curve) and shortest path (blue line) solutions in the Taylor-Green flow %\cite{piro2022optimal}. (b) Snapshot of a learned trajectory (blue curve) and %corresponding policy map (arrows) in presence of virtual obstacles (red cells) for %smart feedback-controlled colloids moving toward a goal (green cell) %\cite{muinos2018}. (c) Point-to-point navigation in a mexican hat potential %(background). Trajectories obtained via Q-learning after 2000 (yellow), 3000 (green) %and 5000 (blue) training episodes are shown in comparison with the exact optimal %trajectory (red) \cite{schneider2019}. (d) Time-evolution of the total reward (blue %curve) and standard deviation (shaded area) for active particles in turbulent flows %(inset) \cite{biferale2019}. See refs for details. 
\label{fig:papers}
\end{center}
\end{figure*}

\begin{figure}
\includegraphics[width=0.49\textwidth]{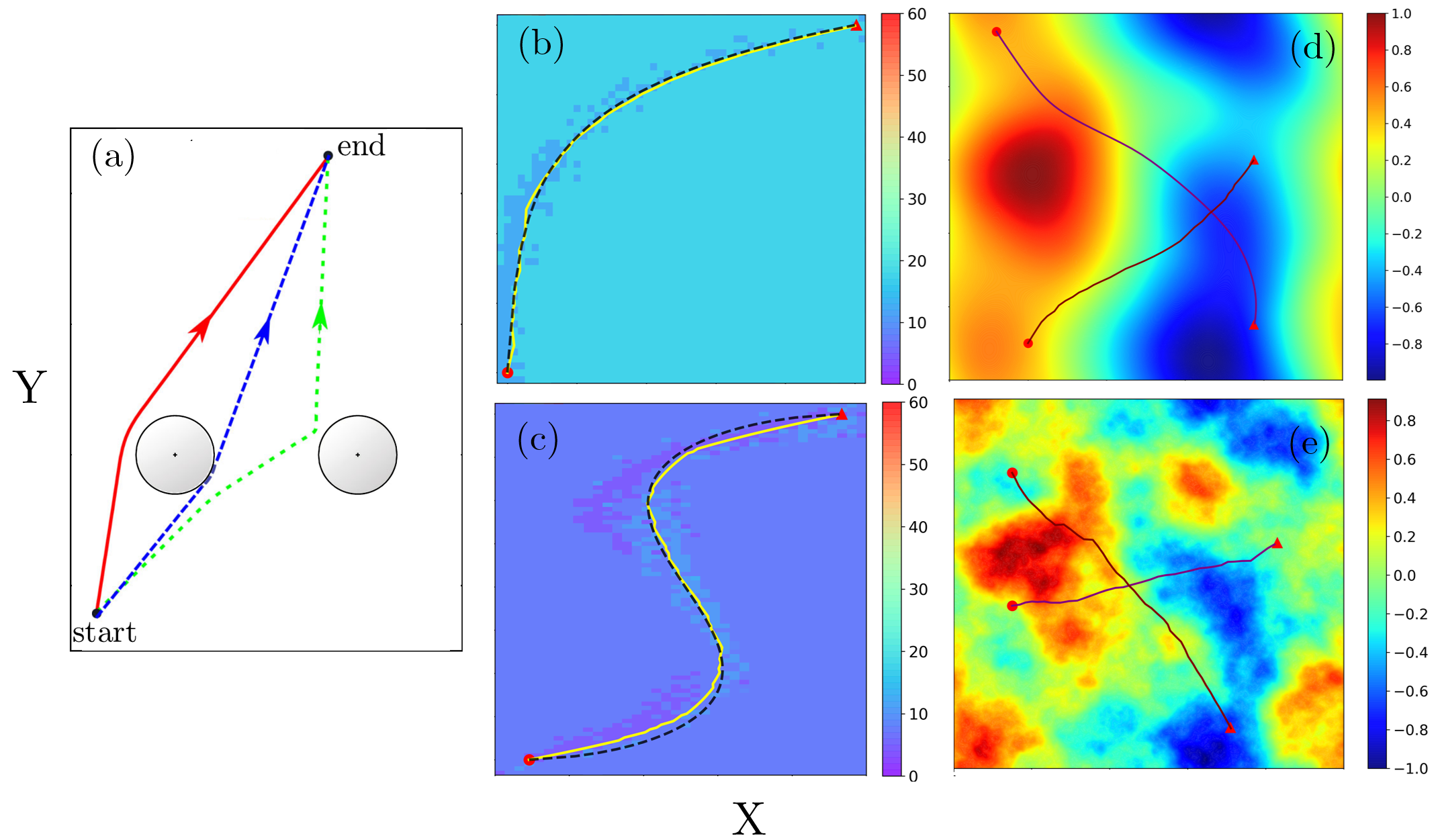}
    \centering
    \caption{(a) Hydrodynamics can qualitatively change the navigation strategy that an active particle needs to follow to reach a target fastest: Optimal trajectory of a dry active particle (blue) and of source dipole microswimmers with source dipole strength $\sigma = -15$ (green) and $\sigma = 7.5$ (red) which interact hydrodynamically with obstacles (grey disks) \cite{daddi2021hydrodynamics}. (b-e) Exact optimal trajectories for active particles (dashed lines) between a given starting and end point (red dots) in a linear force field (b) and in horizontal pipe flow (c) \cite{liebchen2019optimal} in comparison to machine-learned trajectories (yellow lines) \cite{nasiri2022reinforcement}. Background colors show the learned policy map, i.e. the preferred discretized steering “direction” $\psi$ ($\psi\in \{0,..,59\} \cong [0,2\pi)$. (d,e) Lines show learned trajectories in Gaussian random potentials (background colors) \cite{nasiri2022reinforcement}. See refs for more details. }
\label{opttrajs}
%    \vspace{8mm}
%\end{wrapfigure}
\end{figure}

Ernst Zermelo asked in 1931 how a ship needs to steer in a nonuniform wind field to reach its target fastest \cite{zermelo1931navigationsproblem}. Zermelo's problem has later become relevant to a variety of topics ranging from active particles, to fish-like underwater vehicles \cite{yu2018motion}
and unmanned balloon navigation in the stratosphere \cite{bellemare2020autonomous}. 
More generally, in this section, we ask how a self-propelled agent, which is subject to constraints has to steer to optimally (e.g. fastest, cheapest, or safest) reach a target in an environment comprising complex (e.g. turbulent) flow and force fields, motility fields (as relevant for light-powered microswimmers in nonuniform intensity fields), viscosity fields (such as in the case of viscotaxis \cite{liebchen2018viscotaxis}) and complex obstacle landscapes. 
%The required navigation strategy in turn will depend on the environment, the constraints and 
%on the choice for what we optimize \cite{liebchen2019optimal}. 
%Possible candidates are (i) the fastest route, (ii) the cheapest route, i.e. the route leading to minimal fuel requirement, and (iii) the safest route (in the presence of fluctuations). 
% Finally, one can distinguish between fully and partially observable environments, the latter applying to cases where fluctuations are present, as well as between known environments, where the agent learns a navigation strategy which is optimized for a particular environment, 
% and unknown environments, where the agent has to learn transferable strategies which lead to efficient navigation strategies even in environments which the agent has never encountered before. 
The optimal path (trajectory) and the corresponding navigation strategy of a self-propelled agent can be determined e.g. based on 
Pontryagin's principle (for deterministic problems) or Hamilton-Jacobi-Bellman equations (also for stochastic problems) \cite{kirk2004optimal, lewis2017optimal}, geometric approaches \cite{piro2021optimal, piro2022optimal, piro22}, 
modern optimization algorithms \cite{panda2020comprehensive}, but also 
based on (deep) reinforcement learning methods \cite{sutton2018reinforcement,arulkumaran2017brief, franccois2018introduction}, which are particularly useful if the environment is not known (or partially known) 
or for high-dimensional and chaotic environments, where exact solutions are difficult to obtain (Figs. \ref{fig:papers} and \ref{opttrajs}). 
%\begin{figure}
%\begin{center}
%\includegraphics[width=0.4\textwidth]{fluctuations.jpg}
%\end{center}
%\caption{\small Schematic of an active particle aiming to reach a target in minimal time in a globally known %environment which comprises a force/flow field (lines with arrows) and obstacles (grey). 
%When does the optimal path and hence also the navigation strategy of the microswimmer change dramatically in the %presence of thermal fluctuations? 
%A conceivable scenario would be that the microswimmer looses a significant amount of time when colliding with %obstacles, which is provoked by fluctuations, and therefore
%prefers a safer path circumventing narrow ``channels'' between obstacles.
%Lines represent conceivable candidates for optimal trajectories when following the optimal navigation strategy %in the absence (red) and in the presence of fluctuations (green).}
%%\end{wrapfigure}
%\label{schematicendtoend}
%\end{figure}
\begin{figure}
 \includegraphics[width=0.49\textwidth]{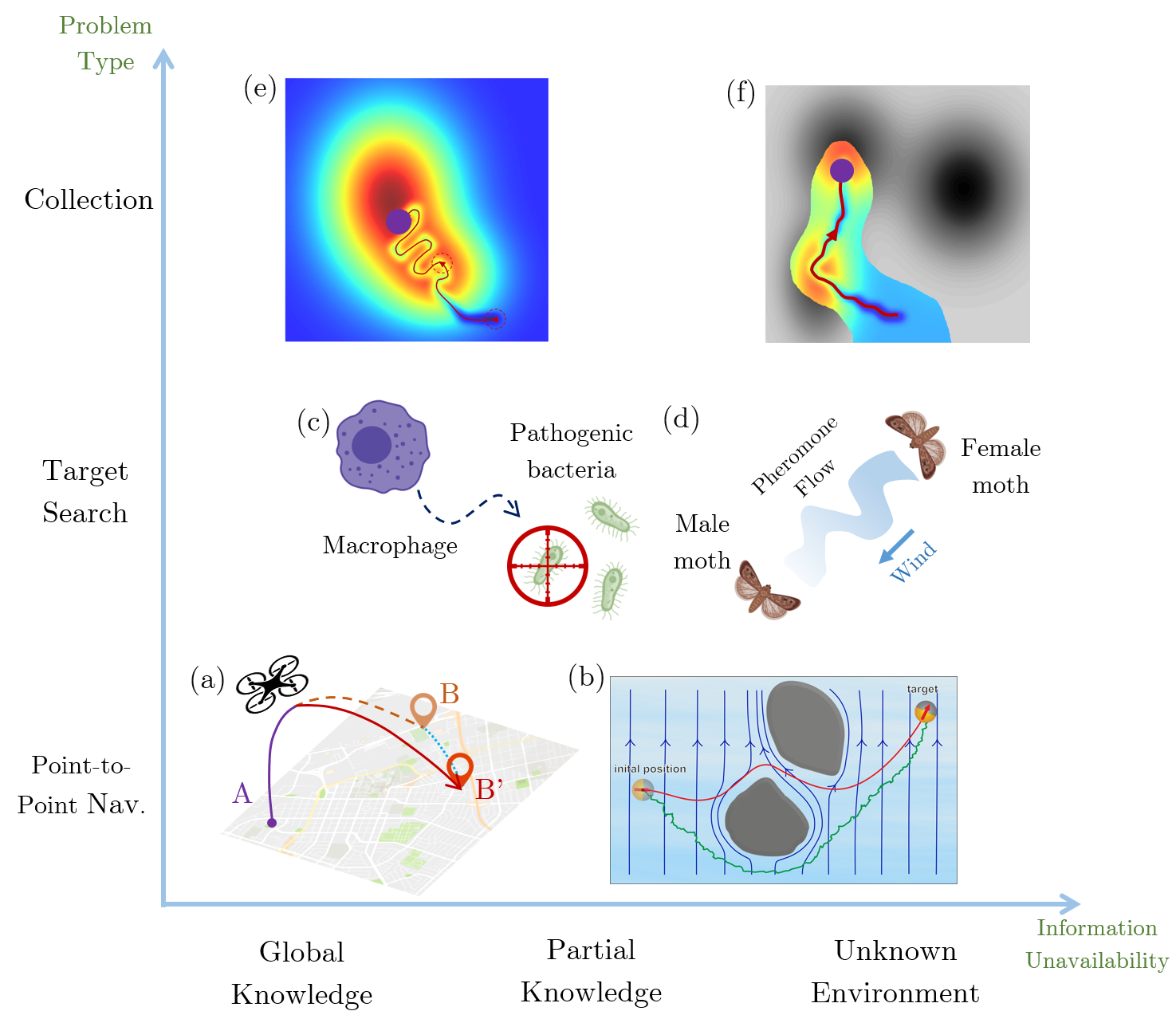}
 \caption{Classification of optimal navigation problems for active particles. 
Examples: Optimal point-to-point navigation in (a) deterministic and (b) fluctuating environments, (c) target search and capture in predator-prey systems, (d) odor search in turbulent streams, optimal collection and localization in (e) known, and (f) unknown setup.}
%A conceivable scenario would be that the microswimmer looses a significant amount of time when %colliding with obstacles, which is provoked by fluctuations, and therefore
%prefers a safer path circumventing narrow ``channels'' between obstacles.
%Lines represent conceivable candidates for optimal trajectories when following the optimal %navigation strategy in the absence (red) and in the presence of fluctuations (green)
 \label{fig:classification}
\end{figure}
% However, if finding the truely optimal trajectories 
% Reinforcement learning can be used to find efficient or even optimal navigation strategies for active particles in complex environments.
\vsc
\underline{Elementary calculation of exact optimal trajectories --}
Consider an overdamped dry active particle (no hydrodynamic interactions) in a 
time-independent and two-dimensional complex environment. 
The equation of motion for the particle position 
$\vec{r}(t) = (x(t), y(t))$ can be compactly written as \cite{liebchen2019optimal}
\begin{equation}
    \Dot{\vec{r}}\,(t) = v_0(\vec r) \hat{n}(t) + \vec{f}(\vec{r}) + \sqrt{2D} \vec{\eta}(t).
    \label{eom1}
\end{equation}
%\begin{wrapfigure}{r}{0.5\textwidth}
Here $\vec f(\vec r)$ represents a general force, flow, and viscosity field and $D$, $\vec \eta$ represent the translational diffusion coefficient and Gaussian white noise of zero mean and unit variance. Let us first focus on the idealized situation where the agent can freely and instantaneously control its self-propulsion direction 
$\hat{n}(t) = (\text{cos}\psi(t), \text{sin}\psi(t))$ but not its speed 
$v_0(\vec r)$, which may depend on space \cite{lozano2016phototaxis,jahanshahi2020realization}.
The goal is to find, for a given starting and end point, the connecting path $\vec r(t)$ (equivalently the steering angle $\psi(t)$), allowing the active particle to reach the target fastest. To solve this problem for the vanishing noise $(D=0)$, we now write the traveling time as a functional of the path $y(x)$ and of $y^{\prime}(x) = d y(x)/d x$ as
\begin{equation}
\small
T\left[y(x), y^{\prime}(x), x\right] = \Int{0}{T}\de t = \int_{x_{A}}^{x_{B}} \4{\mathrm{~d} x}{|\dot x(y(x),y'(x),x)|} 
\normalsize
\label{Tfunc}
\end{equation}
and minimize it by 
solving the Euler-Lagrange equation $\frac{\mathrm{d}}{\mathrm{d} x} \frac{\partial L}{\partial y^{\prime}}-\frac{\partial L}{\partial y}=0$ (boundary value problem) for $L\left(y(x), y^{\prime}(x), x\right)=\frac{1}{|\dot x(y(x),y'(x),x)|}$, where $\dot{x}$ denotes the velocity in x-direction. Using (\ref{eom1}), one obtains the Lagrange function \cite{liebchen2019optimal}
\begin{equation} 
\small
L(x, y(x),y^{\prime}(x))=\frac{\left(1+y^{\prime 2}\right)}{\left|f_{x}+y^{\prime} f_{y} \pm \sqrt{v_{0}^{2}\left(1+y^{\prime 2}\right)-\left(f_{y}-y^{\prime} f_{x}\right)^{2}}\right|},
\label{Lagrangian}
\end{equation} 
where $\vec f = (f_x,f_y)$ and $v_0$ may depend on $x,y$. 
%and where $y'=\de y(x)/\de x$.
% \vsc 
% More general problems (e.g. deterministic time-dependent and 3D problems) can be approached by using 
% Pontryagin's maximum principle \cite{kirk2004optimal}, daddi2021hydrodynamics but do not allow for explicit analytical solutions in most cases.
%\begin{figure}
%\begin{center}
%\includegraphics[width=0.4\textwidth]{stark_Q_learning.png}
%\end{center}
%\caption{\small Point-to-point navigation in a mexican hat potential (background). Optimal trajectories %obtained via Q-learning after 2000 (yellow), 3000 (green) and 5000 (blue) training episodes are shown in %comparison with the exact optimal trajectory (red).
%See \cite{schneider2019} for details.}
%\end{figure}
\vsc\underline{Active particles and ray-optics:} 
The Euler-Lagrange equation, together with Eq. (\ref{Lagrangian}) can be readily solved for constant $\vec f$ and constant $v_0$, yielding $y'(x)=\text{const}$, showing that the shortest path is fastest in any constant field. That is, the active particle steers such that it exactly compensates for the drift due to the environment. In piecewise constant environments, the optimal trajectory is also piecewise constant, yielding Snell's law for active particles which involves a generalized refractive index that can also be negative as for light in meta-materials \cite{shelby2001experimental,smith2004metamaterials}. (See also \cite{ross2021snell}.)
Other exact solutions for the Euler-Lagrange equation can be obtained by exploiting conservation laws 
(symmetries) showing that the shortest path is typically not the fastest in complex environments. In rotating flow fields, active particles sometimes even have to initially swim away from the target to reach it fastest. Note that optimizing other quantities, such as the dissipated power along the path, leads to a different Lagrange function and hence in general also to a different navigation strategy \cite{liebchen2019optimal}.
\vsc\underline{Hydrodynamic interactions --}
%\begin{wrapfigure}[27]{r}{0.35\textwidth}
Instead of dry active particles, ref. \cite{daddi2021hydrodynamics} considers 
microswimmers which hydrodynamically interact with walls and obstacles. One key result was to show that the optimal navigation strategy which microswimmers require can qualitatively 
differ from the one which leads to optimal trajectories for a dry active particle or a macroscopic vehicle in the same environment (Fig.~\ref{opttrajs}a). 
%In the simplest case of a source dipole microswimmer showing hydrodynamic interactions with a remote, infinitely extended, and flat wall, the equations of motion of the swimmer read \cite{spagnolie2012hydrodynamics}: 
%begin{equation}
%v_{x}=\left[v_{0}-\sigma /\left(4 z^{3}\right)\right] \cos \psi, \quad \quad v_{z}=\left(v_{0}-\sigma / z^{3}\right) \sin \psi
%\end{equation}
%and $\quad v_{y}=0$, 
%where $\sigma$ is the source dipole strength and $z$ is the distance to the wall. 
%As above, the task is to find the optimal steering angle $\psi(t)$ (or the corresponding optimal trajectory ${\vec r}(t)$) to minimize the traveling time. 
%The Lagrangian which together with the Euler-Lagrange equations and the boundary conditions (starting and end point) determines the fastest route and then takes the simple form (in units of length $\ell=\left(|\sigma| / v_{0}\right)^{1 / 3}$)
%\begin{equation}
%  \mathcal{L}_{\mathrm{SD}}^{*}=\left(\frac{1}{\left(1-\frac{s}{4 z*^{3}}\right)^{2}}+\frac{z*^{\prime}}{\left(1-\frac{s}{z*^{3}}\right)^{2}}\right)^{1 / 2} 
%\end{equation}
%\noindent where $x^{*}=x / \ell, z^{*}=z / \ell, z^{*^{\prime}}=\partial z^{*} / \partial x^{*} \text { and } s=\operatorname{sgn}(\sigma)$.
%generalized optimal microswimmer navigation problems to account for time-dependent 
%swimmers (which feature both a time-dependent speed and time-dependent hydrodynamic interactions), for which we have used %Pontryagin's minimum principle from optimal control theory. 
%\vsc 
\vsc\underline{Reinforcement Learning --}
In very complex environments (e.g. chaotic or unknown cases), optimal navigation problems can typically not be solved exactly, but methods based on reinforcement learning \cite{sutton2018reinforcement} can still be applied. Some works have used tabular Q-learning, for efficient real-time control of self-thermophoretic active particles (Fig. \ref{fig:papers}b) \cite{muinos2018}, or for learning to navigate optimally inside an environment hosting a Mexican hat potential without brim (Fig. \ref{fig:papers}c) \cite{schneider2019}. 
Other recent works have used actor-critic methods \cite{biferale2019} and also deep reinforcement learning \cite{gunnarson2021learning, nasiri2022reinforcement} to study optimal navigation within increasingly complex environments.
% \vsc
% \\Clearly, as the environment gets too complex, analytical methods as used in [P1,P2] no longer allow us to determine optimal navigation strategies. 
% In addition, as discussed in the introduction, when the environment gets ``strong'', there is no unique receipe on how to generalize classical path planning algorithms 
% such as Dijkstra, A*, etc. \cite{dijkstra1959note,hart1968formal} to reliably determine optimal trajectories. 
In ref. \cite{nasiri2022reinforcement} in particular a deep reinforcement learning-based method has been developed to determine {\it asymptotically optimal trajectories}.
Here, the key challenge was to develop an approach that is capable of finding the global optimum rather than some locally optimal path. The key ``trick'' to meet this challenge was to use a policy gradient-based method that ``understands'' and directly focuses on optimizing the expected total reward (as opposed to the off-policy methods such as Q-learning). To benchmark this method, results were compared to exactly known optimal trajectories (Fig.~\ref{opttrajs}).
%but also to compute optimal trajectories in very complex environments.
%That is, ref. \cite{nasiri2022reinforcement} has shown that it is possible to systematically learn the result of an %optimal control calculation without having to do it.
\\
Ref. \cite{Yang2018} treats the microswimmer navigation problem as a Markov decision problem and minimizes a cost function to solve mazes, assuming global knowledge of the environment (see also  \cite{jin2017chemotaxis} for maze solving with droplets). Very recently, swarms of intelligent colloidal microrobots were also trained for capturing Brownian cargo particles within mazes \cite{xu2021brownian}. Apart from maze solving, recently, a series of studies \cite{Yang2019,yang2020micro,xu2021brownian} have explored microswimmer navigation also in an unknown environment containing obstacles that are locally explored by the microswimmer, using (deep) reinforcement learning. It was found that smart colloids receiving local sensory input were able to navigate around obstacles to reach a target using deep reinforcement learning \cite{Yang2019} and to accomplish complex navigation and localization tasks under time constraints \cite{yang2020micro}. \\
%Ref. \cite{biferale2019}
\noindent As opposed to unknown environments, for the problem of optimal navigation in chaotic (turbulent) environments, one can in principle use optimal control theory (Pontryagin's principle) to determine the exact optimal trajectory. However, this problem is numerically not easily solvable with shooting methods since, in chaotic environments, a tiny variation of the 
initial condition commonly results in a completely different endpoint, and a systematic variation of the initial conditions is not useful and one needs to work with an extended target domain. This difficulty in evaluating the equation representing the exact optimal solution makes the usage of reinforcement learning particularly valuable and accordingly, ref. \cite{biferale2019} has recently used an actor-critic based reinforcement learning approach for Zermelo's problem in turbulent flow fields (Fig. \ref{fig:papers}d). \\ Ref. \cite{gunnarson2021learning} in turn has explored the effects of accounting for environmental cues (such as vorticity, flow velocity, etc.) within the input features of a deep reinforcement learning method and the resulting strategies for a point-to-point navigation task within a turbulent flow. Ref. \cite{alageshan2020machine} in turn used an adversarial reinforcement learning method to train microswimmers for time-efficient point-to-point navigation within statistically homogeneous and isotropic turbulent fluid flows which were able to outperform the naive strategy of always moving in the direction of the target.\\While the swimming direction of synthetic microswimmers can be typically controlled with external fields \cite{bechinger2016active}, many biological microswimmers autonomously change their swimming direction through suitable shape-deformations. 
In line with that, several 
recent works have used reinforcement learning to explore the swimming mechanism of deformable agents 
\cite{tsang2018,liu2021mechanical, qiu2020swimming, behrens2022smart}. Related works have used learning approaches to understand how a swimmer needs to deform to swim as fast as possible \cite{jiao2021learning}, to follow a predetermined path \cite{zou2022gait}, to exhibit chemotaxis \cite{hartl2021microswimmers} or to achieve optimal point-to-point navigation \cite{amoudruz2022independent, zhu2022numerical,zhu2022point}. For example, ref. \cite{zhu2022numerical} considers three-link models of (bionic) fish which receive only their orientation and distance to a (moving) target as input data. They learn generic strategies which are then explored in situations that the swimmer has not encountered throughout training.\\ In another line of work, smart active particles have been trained to exploit underlying turbulent flows to escape local fluid traps \cite{gustavsson2017finding}, reach target regions with high-vorticity \cite{colabrese2018}, or navigate towards the highest altitude achievable \cite{colabrese2017}. Very recently, smart microswimmers equipped with tabular Q-learning were also able to demonstrate efficient navigation strategies (while only having access to local information) within environments hosting various motility fields \cite{monderkamp2022active}.

%-- in a way that allows for generalizations beyond what is accessible to analytical calculations.} The present %project strongly builds on this insight and the developments of \cite{nasiri2022reinforcement}.

\section{Searching and capturing targets}
Another class of optimal navigation problems concerns the quest of how an active agent has to move to efficiently find a target with an unknown location (dynamics).
Here one can distinguish problems (i) where the agent does not receive any information from the target, to which we, therefore, refer as ``silent'' targets, (ii) problems where the target does emit certain information e.g. in the form of odors which is spread by diffusion or advection (Fig. \ref{fig:classification}d), and (iii) problems where the agent is aware of the current location of the target (can "see" the prey) but not of its dynamic (Fig. \ref{fig:classification}c). Other interesting examples occur if the predator has only access to indirect or partial information about the prey (type (ii)), as relevant e.g. for sharks and rays sensing their victims via the created flow fields through lateral line sensors \cite{kalmijn1971electric, gardiner2014flow}, and for chemotactic bacterial predators \cite{chet1971chemical}.
Note that the special case of (iii) where the target does not move corresponds to point-to-point navigation as discussed in the previous section.
\vsc\underline{(i) Finding ``silent'' targets:}
Problems of class (i) have been studied very recently based on 
the development of an algorithm 
generalizing transition-path sampling to active Brownian particle dynamics searching for a target in a complex environment 
\cite{zanovello2021target,zanovello2021optimal}. For self-propelled particles in search of a target located at the center of a circular confining domain, controlled adjustment of parameters such as the self-propulsion velocity and the characteristic rotation time was demonstrated to improve the search efficiency \cite{wang2016target}. Later on, ref.  \cite{wang2017spatial} also studied the role of environmental characteristics (such as spatial heterogeneity) on the target search dynamics and capabilities of self-propelled particles. 
\vsc\underline{(ii) Finding sources:} 
Target search problems have been studied intensively for macroscopic animals searching for odor sources in complex flow fields \cite{celani2014odor,baker2018algorithms,durve2020collective, nguyen2021flow, reddy2022olfactory}. Such flow fields do not only allow for 
pheromone communication among animals but they also efficiently distribute odors over much longer distances than enhanced molecular diffusion would \cite{reddy2022olfactory}.
\noindent However, they also make it difficult to predict the location of the source based on the information which an agent receives from its immediate vicinity.
%The problem of search for odor sources in complex flows has been studied e.g. based on 
A popular strategy for searching with sparse information is infotaxis \cite{vergassola2007infotaxis}, which has been intensively studied in the context of odor search problems for insects. 
This strategy aims at locally maximizing the information gain; i.e. infotactic agents essentially move up the (local) information gain gradient, similarly as chemotactic bacteria move up the concentration gradient. Infotaxis has been recently tested against methods from value iteration (for partially observable Markov-decision problems \cite{spaan2005perseus})
\cite{heinonen2022optimal,rigolli2022alternation} to reinforcement learning
\cite{loisy2022searching,reddy2022sector}.\\
%Despite the existence of all these references, the problem of microswimmers searching for targets of unknown locations in a complex environment has not yet been studied (much) with machine-learning techniques but is expected to be important e.g. for problems like finding (early-stage) cancer cells which typically act as sources for certain molecules \cite{szatrowski1991production}.  
\vsc
\underline{(iii) Catching targets moving in an unknown way:} Predator-prey problems for active Brownian particles have recently been studied using Q-learning \cite{la2014, gerhard2021hunting}. Ref. \cite{borra2022reinforcement} in turn explored the outcome strategies of training adversarial reinforcement learning agents in microswimmer pursuit and evasion tasks. Here, throughout the training, the predator and the prey devised policies to exploit hydrodynamic interactions to out-compete each other with complex sequences of moves and countermoves. Very recently also a deep policy gradient-based method has been demonstrated to be able to qualitatively reproduce the optimal predatory path in chasing a finite size prey at low Reynolds number \cite{zhu2022optimising}.

\vsc \underline{Soaring:}
Another related class of problems, where the target is not necessarily localized and which has been explored with reinforcement learning methods, 
concerns the soaring of birds, unmanned air vehicles, or (other) gliders, i.e the question of how these agents have to navigate to find and navigate thermals within a complex landscape \cite{reddy2016learning,reddy2018glider}. Interestingly, it is still unknown how birds achieve this \cite{reddy2018glider}.

\section{Collection problems} 
How does a prospective microswimmer have to move to efficiently collect targets that are distributed in an unknown way, such as toxins or microplastics? 
This problem, which is closely related to the area sweeping tasks in robotics \cite{ahmadi2005continuous,ahmadi2006multi,shah2020deep}, has not yet been studied much in the active matter literature (Fig \ref{fig:classification}). 
%Very recently [Grauer work,Edwin work?] 
\\Existing works largely focus on stochastic search strategies. Several influential works have reported observations of such strategies in the forms of Levy walks (step size randomly drawn from a fat-tailed distribution) in the foraging of albatrosses \cite{viswanathan1996levy}, marine predators \cite{sims2008scaling}, bumble bees and T cells \cite{zaburdaev2015levy}.
Ref. \cite{volpe2017topography} explores using Levy walking active particles to collect 
(nonregenerative) sparse targets. While in homogeneous environments a certain combination of diffusive and ballistic motions is believed to be optimal, this work finds that as the environment gets increasingly complex due to the presence of 
barriers, strategies that are more diffusive tend to lead to better target collection rates. 
Similarly, run-and-tumble walkers searching for a single target have been studied in ref. \cite{rupprecht2016optimal}.
\vsc
%\section{Other problems}
%Ref. \cite{durve2020learning} uses reinforcement learning to allow particles to learn aligning their %orientation with each other and to form a flock. (This can be viewed as a navigation problem for the %orientational degree of freedom of the particles).
%Refs. \cite{dou2019autonomous,hartl2021microswimmers} consider shape-deformable swimmers and discuss %strategies on how to 
%program their deformation pattern to realize a desired response to their environment (such as %chemotaxis). Ref. \cite{hartl2021microswimmers} uses artificial neural networks and genetic algorithms to %learn chemotaxis, whereas \cite{dou2019autonomous} uses an evolutionary algorithm (the so-called 
%``covariance matrix adaptation evolution strategy'' 
%to minimize the relevant cost function. \textcolor{blue}{Please double check information in this %sentence.}\textcolor{purple}{@Benno: Thought we wont cover this?Also I think we should discard this %section}
%
%Motivated by the swimming of ``active'' plankton
%which in reality can rotate themselves and swim into a desired direction 
%and are able to sense and react to ambient information including the motion of ambient flow 
%\cite{qiu2020swimming}, ref. 
%\cite{qiu2020swimming} uses Q-learning to find strategies for finding efficient navigation strategies for %moving in the upwards flow direction in a two-dimensional vortex flow pattern. 
\textbf{Open questions, challenges and perspectives.\\}
\underline{How good are machine-learned results? --}
When learning the result of a navigation problem (or of another complex control problem), it often remains unclear how close the resulting trajectory or navigation strategy is to the real optimum. 
Of course, convergence of the reward does not guarantee optimality: The reward can converge to an arbitrary local optimum \cite{sutton2018reinforcement} and 
even if it converges to the global optimum it is often unclear if this optimum is representative for the (asymptotic) physical optimum, or just optimal within the given reward definition, discretization, hyperparameter choice, and the chosen learning algorithm. 
Accordingly, one major open challenge is to develop reinforcement learning approaches that can reproduce exact results and which can be used to go beyond those in \cite{nasiri2022reinforcement}.
%Besides developing ``better'' reinforcement learning approaches for optimal navigation problems it is %therefore important to also 
%develop smart (human-designed) strategies which can be used to benchmark results obtained by machine %learning. 
%For target detection problems in insects for example, strategies like infotaxis %\cite{vergassola2007infotaxis} and variants thereof have been used as benchmarks. 
%
\vsc
\underline{Fluctuations --}
While we expect that fluctuations can qualitatively change (Fig. \ref{fig:classification}b) the required navigation strategy to optimally reach a target (Fig. \ref{fig:papers}a) \cite{piro2021optimal}, as e.g. in the cliff walking problem \cite{van2009theoretical}, they have little effect on the required navigation strategies in other problems.
Accordingly, it would be important to systematically understand and formulate criteria for when fluctuations lead to strong quantitative or even qualitative changes in the required navigation strategy. 
As for the development of reinforcement learning algorithms, fluctuations lead to environments that are only partially observable, hence making the outcome of decisions (actions) made by the agents not accurately predictable. In certain problems, this unpredictability can challenge the robustness of training, which highlights the need for novel methods based on deep reinforcement learning (such as trust region methods \cite{schulman2015trust, schulman2017proximal}) which are capable of maintaining robust learning in volatile setups.
\vsc
\underline{Transferability \& unknown environments --}
While microorganisms require strategies to navigate and find food in environments that they have never encountered before and which may change over time in an unpredictable way, many navigation problems for active particles so far hinge on a fixed (or deterministic dynamic) environment.
An early work that addresses optimal navigation of ``colloidal robots'' in unknown environments based on deep reinforcement learning is \cite{Yang2019}.
Developing powerful methods to allow determining transferable navigation strategies in the future will likely require methods from model-based reinforcement learning \cite{kaiser2019model, moerland2020model} and more importantly world models \cite{schrittwieser2020mastering} where the agents strive to learn a model (representation) of the environment (which here can even be translated to learning the physics of the setup) and use this representation of the environment for planning its future actions.
\vsc
\underline{Recent developments in machine learning --}
We are currently witnessing a rapid advancement in the development of new reinforcement learning methods. Accordingly, it is not 
surprising that some of the most powerful methods have not yet been applied to active matter and related optimal navigation problems. A very promising line of study that has been recently applied to famous games such as Chess, Go, and Shogi is the introduction of reinforcement learning algorithms with integrated planning such as the AlphaGo and AlphaZero \cite{silver2016mastering, silver2017mastering,silver2018general}. We believe, given the robustness of their planning phase (thanks to a built-in Monte Carlo tree search of possible future outcomes), these methods can be very useful in tasks requiring high degrees of accuracy and confidence in the optimality of the learned strategies.
\vsc
\underline{Cross-interactions and communication rules --}
Recently, several works have focused on motile agents learning collective behaviors 
\cite{charlesworth2019intrinsically} such as flocking from simple low-level principles and incentive designs with reinforcement learning techniques \cite{durve2020learning, hornischer2022modeling}. 
A related line of study concerns the application of multi-agent reinforcement learning \cite{zhang2021multi} to microswimmer problems. One can imagine complex tasks such as localizing cancer cells or collecting microplastics while having low environmental awareness (Fig. \ref{fig:classification}f), which would require multiple smart microswimmers to cooperate and share their gathered knowledge of the host environment to guarantee success.

%could enable learning and developing collectively optimal navigation and communication strategies for groups of microswimmers, allowing them to share their gathered knowledge of the host environment for achieving the goal.

\bibliographystyle{eplbib}
%\bibliographystyle{unsrt}

% \begin{thebibliography}{}
% \expandafter\ifx\csname url\endcsname\relax\def\url#1{\texttt{#1}}\fi
% 
% \bibitem{bronn1862klassen}\Name{Bronn, H.G.}\REVIEW{Klassen und Ordnungen des Thier-Reichs (CF Winter)}{}{1862}{}
% 
% \end{thebibliography}
\bibliography{optnavShort.bib}

%\printbibliography
\end{document}